\documentclass [prb,preprint,superscriptaddress,floatfix,amsmath,amssymb] {revtex4-1}
\usepackage{amsmath}
\usepackage{graphicx}
\usepackage{hyperref}
\usepackage{color}
\usepackage{amssymb}
\usepackage{bm}
\newcommand{\beq} {\begin{equation}}
\newcommand{\eeq} {\end{equation}}
\newcommand{\bea} {\begin{eqnarray}}
\newcommand{\eea} {\end{eqnarray}}
\newcommand{\be} {\begin{equation}}
\newcommand{\ee} {\end{equation}}

\newcommand{\bo}{\bar \omega}
\newcommand{\bO}{\bar\Omega}
\newcommand{\bL}{\bar\Lambda}
\newcommand{\bP}{\bar\Phi}
\newcommand{\bS}{\bar\Sigma}
\newcommand{\BP}{\bar\Phi}
\newcommand{\BS}{\bar\Sigma}

\newcommand{\bD}{\bar \Delta}
\newcommand{\BD}{\bar \Delta}
\newcommand{\bT}{\bar T}
\DeclareMathOperator{\sgn}{sgn}

\definecolor{mycolor}{RGB}{254,0,0} 

\begin{document}
\title {Pairing by a dynamical interaction in a metal}
\author{Andrey V. Chubukov}
\affiliation{School of Physics and Astronomy and William I. Fine Theoretical Physics Institute,
University of Minnesota, Minneapolis, MN 55455, USA}
\author{Artem Abanov}
\affiliation{Department of Physics, Texas A\&M University, College Station,  USA}
\date{\today}
\begin{abstract}
 We  consider  pairing of itinerant fermions in a metal near a quantum-critical point (QCP) towards
  some  form of particle-hole order (nematic, spin-density-wave, charge-density-wave, etc).  At a QCP, the dominant interaction between fermions comes from exchanging  massless   fluctuations of a critical order parameter. At low energies, this physics can be described by an effective model with the dynamical electron-electron interaction
 $V(\Omega_m) \propto 1/|\Omega_m|^\gamma$, up to some upper cutoff $\Lambda$.  The case $\gamma =0$ corresponds to BCS theory, and can be solved by summing up geometric series of Cooper logarithms. We show that for a finite $\gamma$, the pairing problem is  still marginal (i.e., perturbation series are logarithmic), but one needs
 to go beyond logarithmic approximation to find the pairing instability.   We discuss specifics of the pairing at $\gamma >0$ in some  detail and also analyze  the marginal case
  $\gamma = 0+$, when  $V(\Omega_m) = \lambda \log{(\Lambda/|\Omega_m|)}$.   We show that in this case the summation of Cooper logarithms does  yield the pairing instability at $\lambda \log^2{(\Lambda/T_c)} = O(1)$, but  the logarithmic
   series are not geometrical.   We reformulate the pairing problem in terms of a renormalization group (RG)  flow of the coupling, and show that the RG equation is different in the cases $\gamma=0$, $\gamma = 0+$, and $\gamma >0$.

 \end{abstract}
\maketitle

\section{ Preface}

It is our great pleasure to present this paper for the special issue of JETP devoted to 90th birthday of Igor Ekhielevich Dzyaloshinskii.  His works on  correlated electron systems are of highest scientific quality. He made seminal contributions to quantum magnetism, superconductivity, Fermi-liquid theory, and to other subfields of modern condensed matter physics.  In a number of works, including the famous ones  on the interplay between d-wave superconductivity and  antiferromagnetism at the  beginning of   ``high -$T_c$ era'' (Refs. ~\cite{dzyal,dzyal_1,dzyal_2}),  Igor Ekhielevich used the renormalization group technique to obtain the flow of the couplings upon integrating out fermions with higher energies.  In this study, we apply the same RG technique to analyze the pairing instability in systems with critical dynamical pairing interaction. We consider this as a  natural extension of his works.
  Happy birthday, Igor Ekhielevich, and the very best wishes.

\section{ Introduction.}

 The pairing
 near a quantum-critical point (QCP) in a metal and its interplay with non-Fermi-liquid behavior in the normal state, is a fascinating subject, which attracted  substantial attention in the correlated electron community after the discovery of superconductivity (SC) in
  the cuprates, Fe-based systems, heavy-fermion materials, organic materials,  and, most recently, twisted bilayer graphene~\cite{scal,scal2,book1,review4,Norman_review,matsuda,Fernandes_2016,mack,Rice,review3,Sachdev2018,
  Coleman_book,Cao2018}.
   Itinerant QC models, analyzed in recent years, include  models of fermions in spatial dimensions $D \leq 3$,
    various two-dimensional  models near zero-momentum spin and charge  nematic instabilities, and instabilities towards spin and charge density-wave order  with either real or imaginary (current) order parameter,  2D fermions at a half-filled Landau level, Sachdev-Ye-Kitaev (SYK) and SYK-Yukawa models,  strong coupling limit of electron-phonon superconductivity, and even color superconductivity of quarks, mediated by gluon exchange.  These problems have also been studied analytically and using various numerical techniques.
    We refer a reader to Ref. \cite{paper_1}, where the extensive list of references to these works has been presented.

From theory perspective,  pairing near a QCP is a fundamentally novel phenomenon
 because an effective dynamic electron-electron interaction, $V(q, \Omega)$,
 mediated by a critical collective boson, which condenses at a QCP, provides a strong attraction in one or more pairing channels and, at the same time,
  gives rise to a non-Fermi liquid (NFL) behavior in the normal state.
  The two tendencies compete with each other: fermionic incoherence, associated with the NFL behavior, destroys the Cooper logarithm and reduces the tendency to pairing,  while an opening of a SC gap eliminates the scattering at  low energies  and reduces the tendency to a NFL.
  To find the winner of this competition (SC or NFL), one needs to analyze  the set of integral equations for the fermionic self-energy, $\BS ({\bf k}, \omega)$, and the gap function, $\Delta ({\bf k}, \omega)$,
    for fermions with momentum/frequency $({\bf k}, \omega)$ and $(-{\bf k}, - \omega)$.

  We consider the subset of models, in which collective bosons are slow modes compared to dressed fermions, for one reason or the other.
  In this situation, which bears parallels  with Eliashberg theory for
    electron-phonon interaction~\cite{eliashberg},  the self-energy and the pairing vertex can be approximated by their values on the Fermi surface (FS) and computed within the one-loop approximation.
        The self-energy on the FS, $\BS ({\bf k}, \omega)$,  is invariant under rotations from the point group of the underlying lattice. The rotational symmetry of the gap function  $\Delta ({\bf k}_F, \omega)$ and  the relation between the  phases of  $\Delta ({\bf k}_F, \omega)$ on different FS's in multi-band systems
      are model specific. E.g.,  near an antiferromagnetic  QCP in a system with a
       single FS,  the strongest attraction is in the $d-$wave channel.
        In each particular case,  one has to project the pairing interaction into the irreducible channels, find the strongest one,  and solve for the pairing vertex for a given pairing symmetry.

Away from a QCP, the effective  $V(\Omega)$ tends to a finite value at $\Omega =0$. In this situation, the fermionic self-energy has a FL form at the smallest frequencies, and the equation for $\Delta (\omega)$ is similar to that in a conventional Eliashberg theory for phonon-mediated superconductivity;  the only qualitative distinction  for electronically-mediated pairing is that $V(\Omega)$ by itself changes below $T_c$ due to feedback from fermionic pairing on collective modes. At a QCP, the situation is qualitatively different because the effective interaction $V(\Omega)$, mediated by a critical massless boson,  is  a singular function of frequency.   Quite generally, such  interaction behaves at the smallest $\Omega_m$ on the Matsubara axis as  $V(\Omega_m) \propto 1/|\Omega_m|^\gamma$, where $\gamma > 0$ is some exponent.  (Fig. \ref{fig1}).  This holds at
 frequencies below some upper cutoff $\Lambda$.  At larger $\Omega_m > \Lambda$, the interaction drops even further, and can be safely neglected.

 In these notations,  BCS  pairing  corresponds to  $\gamma=0$.  The superconducting transition temperature $T_c$ inthe BCS case can be most straightforwardly obtained by computing the pairing susceptibility $\chi_{pp}$ in the order-by-order expansion in the coupling
 and identifying the temperature at which it diverges.  The series contain the  powers of the product of the Cooper logarithm
  $\log{\Lambda/T_c}$ and the dimensionless coupling  $\lambda$, defined such that $1 + \lambda$ is the ratio of the renormalized and the bare electron masses.  The series
 are geometrical, i.e., $\chi_{pp} = \chi_0 (1 +  \lambda \log{\Lambda/T_c} + (\lambda \log{\Lambda/T_c})^2 + ...) = \chi_0/(1-\lambda \log{\Lambda/T_c})$.   We will see that for $\gamma >0$,  the
    pairing kernel  still has a  marginal $1/|\omega|$ form, and, as a result, the series for $\chi_{pp}$  still contain logarithms. However, contrary to the BCS case, where $1/|\omega|$  comes from the fermionic propagator, the marginal  exponent $-1$ for $\gamma >0$ is the sum of the exponent $\gamma$ from the interaction and $1-\gamma$ from the fermionic self-energy (see below). As a result,
     the logarithmic series are not geometric, and we will see that $\chi_{pp}$ does not diverge down to $T=0$.   We show that the pairing instability still exists, however one needs to go beyond the logarithmic approximation to obtain it.

  We show that there exists a special case, which falls in between $\gamma =0$ and $\gamma >0$.  This is the limit $\gamma \to 0+$, when $V(\Omega_m) \propto \log{\Lambda/|\Omega_m|}$.  In this limit, the  pairing instability still can be detected by summing up the series of logarithms for $\chi_{pp}$. However, the series are  not geometric, and some extra efforts are needed to sum them up to find the form of $\chi_{pp}$ and detect the pairing instability at $T \sim \Lambda e^{-\pi/(2\sqrt{\lambda})}$.

 The model with the interaction $V(\Omega_m) \propto 1/|\Omega_m|^\gamma$ displays a very rich physics,  and our group has studied it over the last few years.  This physics is particularly interesting for $\gamma \leq 2$, where a wide range of the pseudogap (preformed pair) behavior emerges, and for $\gamma >2$, when the new, non-superconducting ground state emerges~\cite{paper_4}. In this communication, we will not discuss these values of $\gamma$ but instead focus on small $\gamma$ and discuss in some detail the difference between BCS pairing at $\gamma =0$ and the pairing  at $\gamma = 0+$ and at  $\gamma>0$. We derive the Eliashberg equation for the pairing vertex
  and analyze it within logarithmic perturbation theory and beyond it.  We then convert the integral Eliashberg equation into the approximate differential equation and obtain and solve the corresponding renormalization group (RG) equation.

  The pairing problem at small $\gamma$  attracted a lot of attention in the last few years from various physics sub-communities of physicists~\cite{senthil,max_last,raghu_15,raghu,*raghu2,*raghu3,*raghu4,*raghu5,Wang_H_17,Wang_H_18,Fitzpatrick_15,
  wang,Wu_19,son,son2,wilczek,pisarski,rischke,torroba_last} .
  The pairing interaction $V(\Omega) \sim 1/|\Omega|^\gamma$ with $\gamma \ll 1$ emerges for fermions near  a generic particle-hole instability in a weakly anisotropic 3D system (more explicitly,  in dimension $D=3-\epsilon$, where $\epsilon \sim \gamma$, Refs. ~\cite{senthil,max_last,raghu_15} ). A similar gap equation with small $\gamma$ holds for the pairing in graphene~\cite{khveshchenko}.  The model with $\gamma = 0+$ describes the pairing in 3D systems and
  color superconductivity of quarks due to gluon exchange~\cite{son,wilczek,pisarski,rischke}.   The $\gamma = 0+$ model  yields a marginal Fermi liquid form of the fermionic self-energy in the normal state and was argued~\cite{chandra_first,*Littlewood_92,chandra_last}
    to be relevant to pairing in the cuprates and, possibly, in Fe-based superconductors.

 The structure of the paper is at follows. In the next Section we present the set of coupled Eliashberg equations  for the pairing vertex $\Phi (\omega_m)$ and the fermionic self-energy $\BS (\omega_m)$ and combine them into the equation for the gap function $\Delta (\omega_m)$.
In Sec. \ref{sec:perturbation}
  we analyze the structure of the logarithmic perturbation theory for $\gamma = 0+$ and $\gamma >0$.
  In Sec. \ref{sec:finite_gamma} we go beyond perturbation theory  and
  re-express the integral Eliashberg equation as an approximate differential equation for the pairing vertex and solve it.   We show that for $\gamma = 0+$, the solution
   coincides
   with the result of summation of logarithmic series.  For $\gamma >0$, we show that
   the system does become unstable towards pairing if the interaction exceeds a certain threshold.
   In Sec. \ref{sec:RG} we analyze the pairing at small $\gamma$ from RG perspective and reproduce the results of the previous Section. We present our conclusions in Sec. \ref{sec:conclusions}.

\section{The model}
\label{sec:model}

 We  consider  itinerant fermions  at the onset
of a  long-range particle-hole order in either spin or charge channel.  At a critical point, the propagator of a soft boson
 becomes massless and
 mediates singular interaction between fermions. A series of earlier works on spin-density-wave order, charge-density-wave order, Ising-nematic order, etc (see Ref.\cite{paper_1} for references) have found
  that this interaction is attractive in at least one pairing channel. We take this as in input and project boson-mediated interaction into the channel with the strongest attraction.
  As we said, at small frequencies, the interaction scales as
$V(\Omega_m) \propto 1/|\Omega_m|^\gamma$.
    We incorporate dimension-full factors, like the density of states, into $V(\Omega_m)$ and  treat it as dimensionless.
   We assume that the power-law form holds
   up to the scale $\Lambda$, and set $V(\Omega_m) =0$ above this scale.
   To keep $V(\Omega_m)$ continuous at $\Lambda$ and also to transform gradually between $\gamma =0+$ and
    $\gamma >0$, we use the following form for $V(\Omega_m)$ (Fig.\ref{fig1}):
    \begin{figure}
	\begin{center}
\includegraphics[width=0.6\columnwidth]{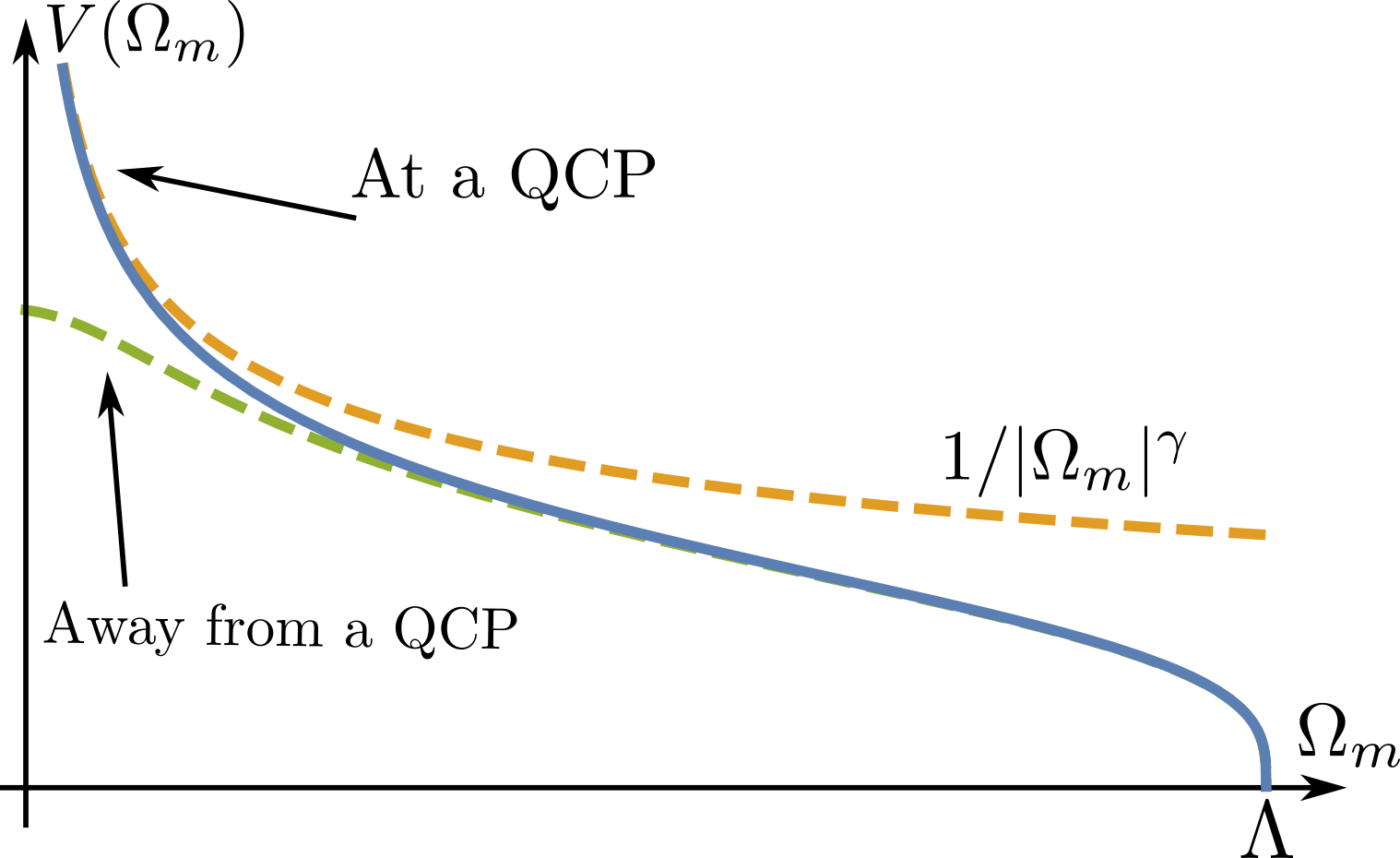}
		\caption{
 Frequency dependence of the effective interaction $V(\bO_m)$, mediated by a soft boson, at $T=0$.
 Away from a QCP, $V(\bO_m)$ tends to a finite value at $\bO_m =0$.  Right at a QCP, the boson becomes massless, and $V(\bO_m)$ diverges as $1/|\bO_m|^\gamma$.}
\label{fig1}
	\end{center}
\end{figure}
      \beq
   V(\Omega_m) = \frac{\lambda}{\gamma} \left(\frac{{\bar g}}{|\Omega_m|}\right)^\gamma \left(1 - \left(\frac{|\Omega_m|}{\Lambda}\right)^\gamma\right)
   \label{a_1}
   \eeq
  Here $\lambda$ is a dimensionless coupling, and ${\bar g}$ has units of energy.
    We assume that $\lambda \ll 1$, but the ratio $\lambda/\gamma$ can be arbitrary.
At $\gamma = 0+$, the last term in (\ref{a_1}) can be expanded in $\gamma$ and yields the logarithmic interaction
  \beq
   V(\Omega_m) = \lambda \log{\frac{\Lambda}{|\Omega_m|}}
   \label{a_2}
   \eeq
The conventional BCS/Eliashberg case in this notations is obtained by extending (\ref{a_1}) to the case when a
    pairing boson has a finite mass.  In this situation, $\Omega_m$ is replaced by a constant below a certain scale.  In BCS theory, this scale is assumed to be comparable to  $\Lambda$, such that $V(\Omega_m) \approx \lambda_{BCS}$ up to $|\Omega_m| \sim \Lambda$, and zero at larger $|\Omega_m|$.

   Below we measure all quantities with the dimension of  energy, i.e., $T, \Omega_m, \Lambda$, in units of ${\bar g}$, i.e., introduce ${\bar T}=T/{\bar g}, {\bar \Omega}_m = {\Omega_m}/{\bar g}$, and ${\bar \Lambda} = \Lambda/{\bar g}$.  Throughout this paper we assume that $\bL \gg 1$.

Earlier works on the pairing mediated by a soft collective boson found that
  a
  boson is overdamped due to Landau damping into a particle-hole pair  and can be treated as slow mode compared to a fermion, i.e., at the same momentum $q$, a typical fermionic frequency is much larger than  a typical bosonic frequency. This is the same property that
  justified  Eliashberg
     theory of phonon-mediated superconductivity.  By analogy,  the theory of electronic superconductivity, mediated by  soft collective bosonic excitations in spin or charge channel, is  often referred to as  Eliashberg theory, and  we will be using this convention.

In Eliashberg  theory,
 one can explicitly integrate first over the momentum component perpendicular to the Fermi surface and then over the component(s) along the Fermi surface, and reduce the
   pairing problem  to a set of coupled integral equations for frequency dependent self-energy $\BS (\omega_m)$
   and the pairing vertex $\Phi (\omega_m)$.

  At $T=0$, the coupled Eliashberg equations are
    \bea \label{eq:gapeq}
    \bP (\bo_m) &=&
     \frac{1}{2} \int d \bo'_{m} \frac{\Phi (\bo'_{m})}{\sqrt{{\tilde \bS}^2 (\bo'_{m}) +\bP^2 (\bo'_{m})}} { V}(|\bo_m - \bo'_{m}|), \nonumber \\
     {\tilde \bS} (\bo_m) &=& \bo_m
   +  \frac{1}{2}  \int d \bo'_{m}  \frac{{\tilde \bS}(\bo'_m)}{\sqrt{{\tilde \bS}^2 (\bo'_{m})  +\bP^2 (\bo'_{m})}} {V}(|\bo_m - \bo'_{m}|)
 \eea
 where ${\tilde \bS}(\bo_{m}) = \bo_m + \bS (\bo_m)$.
   In these equations, both $\bS (\omega_{m})$ and $\bP (\omega_{m})$ are real functions.
  Observe that we define $\bS (\omega_m)$ with the overall plus sign and  without the overall factor of $i$.
  In the normal state ($\bP \equiv 0$), we have at $\gamma >0$, and $|\bo_m| \ll 1$
  \beq
  {\tilde \BS} (\bo_m) = \bo_m + \frac{|\bo_m|^{1-\gamma}}{1-\gamma} \sgn{\bo_m},
\label{ss_111_a}
\eeq
At $\gamma =0+$,
  \beq
  {\tilde \BS} (\omega_m) = \bo_m \left(1+ \lambda \log{\frac{\bL}{|\bo_m|}}\right),
\label{ss_111_a_1}
\eeq

The  superconducting gap function $\BD (\bo_m)$ is defined as the ratio
\beq
 \BD (\bo_m) = \bo_m  \frac{\BP (\bo_m)}{{\tilde \BS} (\bo_m)} = \frac{\BP (\bo_m)}{1 + \BS (\bo_m)/\bo_m}.
  \label{ss_1}
  \eeq
   The equation for $\BD (\omega_{m})$ is readily obtained from (\ref{eq:gapeq}):
   \beq
   \BD (\bo_m) = \frac{1}{2}  \int d \bo'_{m}  \frac{\BD (\bo'_{m}) - \BD (\bo_m) \frac{\bo'_{m}}{\bo_m}}{\sqrt{(\bo'_{m})^2 +\BD^2 (\bo'_{m})}}
    ~\frac{1}{|\bo_m - \bo'_{m}|^\gamma}.
     \label{ss_11}
  \eeq
   This equation contains a single function $\BD (\bo_{m})$, but  for the cost that $\BD (\bo_m)$ appears also in the r.h.s., which makes Eq. (\ref{ss_11}) less convenient for the analysis than Eqs. (\ref{eq:gapeq}).

For a generic $\gamma >0$, the coupled equations (\ref{eq:gapeq})  for $\BP$ and ${\tilde \BS}$  describe the interplay between the two competing tendencies -- one towards superconductivity, specified by $\BP$,  and the other towards incoherent NFL normal-state behavior, specified by ${\tilde \BS}$.  The competition between the two tendencies is encoded in the fact that ${\tilde \BS}$ appears in the denominator of the equation for $\BP$ and $\BP$ appears in the denominator of the equation for ${\tilde \BS}$.
 In more physical terms, a self-energy ${\tilde \BS}$ is an obstacle to Cooper pairing, while when $\BP$  is non-zero, it reduces the strength of the self-energy and renders fermionic coherence.

The full set of the equations for electron-mediated pairing generally must contain the third equation, describing the feedback from the pairing on the bosonic propagator. This feedback is small in the case  of electron-phonon interaction, but
 is  generally not small when the pairing is mediated by a collective mode because the dispersion of a collective mode changes qualitatively below $T_c$ (Refs. \cite{acs,finger_2001}).
In this work, we only consider the computation of $T_c$ and will not
  discuss system behavior below $T_c$.  For this computation,  it is sufficient to restrict with the two equations (\ref{eq:gapeq}). Moreover,  we can (i) treat  $\BP (\bo_m)$ as
 infinitesimally small and  neglect it in the denominator of Eq. (\ref{eq:gapeq}), and (ii) use Eqs. (\ref{ss_111_a})  or
 (\ref{ss_111_a_1}) for ${\tilde \BS}$.
 For a small, but finite $\gamma >0$, the linearized equation for the pairing vertex is
 \beq
   \bP (\bo_m) = \frac{1}{2} \int d \bo'_m  \frac{\bP (\bo'_{m})}{|\bo'_m|^{1-\gamma}|\bo_m - \bo'_{m}|^\gamma (1 + (\gamma/\lambda)|\bo'_m|^\gamma)} \left(1 - \left(\frac{|\bo_m - \bo'_{m}|}{\bL}\right)^\gamma\right)
     \label{ss_111_l}
  \eeq
    Observe that the overall coupling is just a number, equal to one, because the factor $\lambda/\gamma$  in $V(\bO_m)$ cancels out with
  the same factor in the self-energy.   This factor is still present in the denominator, but in the term, which
   contains $|\bo'_m|^\gamma$ and becomes irrelevant  at the smallest frequencies

   At  $\gamma =0+$, the linearized equation for $\bP (\bo_m)$ is
\beq
   \bP (\bo_m) = \frac{\lambda}{2} \int d \bo'_m \frac{\bP(\bo'_{m})}{|\bo'_m|} \frac{\log{\frac{\bL}{|\bo_m - \bo'_{m}|}}}{1 + \lambda \log{\frac{\bL}{|\bo_m - \bo'_{m}|}}}
     \label{ss_111_l_1}
  \eeq

 \section{Summing up the logarithms}
 \label{sec:perturbation}

 As a first try, we analyze the linearized equation for the pairing vertex  perturbatively, by adding
  up a trial, infinitesimally small $\bP_0$ to the r.h.s of Eqs. (\ref{ss_111_l}) and (\ref{ss_111_l_1}) and computing the pairing susceptibility perturbatively, order-by-order.  Such approach has been commonly used for BCS pairing. The perturbation series there contain singular Cooper logarithms.
  The logarithmic singularity can be cut either by a finite $T$ or at $T=0$, by a finite total incoming bosonic frequency, $\bO_{tot}$.  For consistency  with other cases, we set $T=0$ and keep $\bO_{tot}$ finite.  The result of order-by-order analysis for a BCS pairing is well known:
\beq
\BP (\bO_{tot}) = \BP_0 \left(1 + \lambda_{BCS} \log{\frac{\bL}{|{\bO}_{tot}|}} + \lambda^2_{BCS} \log^2{\frac{\bL}{|{\bO}_{tot}|}} + ...\right) = \frac{\BP_0}{1 - \lambda_{BCS} \log{\frac{\bL}{|\bO_{tot}|}}}
\label{su_1}
\eeq
The ratio $\BP (\bO_{tot})/\BP_0$ (the pairing susceptibility) diverges at $|\bO_{tot}| = \bL e^{-1/\lambda_{BCS}}$ and becomes negative at smaller $|\bO_{tot}|$, indicating that the normal state is unstable towards pairing. (In a more accurate description, the pole in $\BP (\bO_{tot})$ moves from the lower to the upper half-plane of  complex frequency~\cite{agd}).

We now do the same calculation for $\gamma >0$.  The kernel in Eq. (\ref{ss_111_l}) is still marginal at $\bo'_m, \bo_m \ll 1$, but, as we said,  now the scaling dimension $-1$ is  the sum of $-\gamma$, coming from the interaction, and $-1+\gamma$, coming from the self-energy.  This implies that perturbation theory still contains logarithms, but in distinction to BCS, each logarithmical integral $\int d \bo'_m/|\bo'_m|$  runs between  the upper cutoff at $|\bo'_m| \sim \bo_{max} = (\lambda/\gamma)^{1/\gamma}$  and
   the lower cutoff at $|\bo'_m| \sim |\bo_m|$. Because the lower cutoff is finite, we can set $\bO_{tot} =0$.
   Summing up the logarithmical series, we then obtain~\cite{acf}
\beq
\BP (\bo_m) = \BP_0 \left(1 +  \log{\frac{\bo_{max} }{|\bo_m|}} +  \frac{1}{2} \log^2{\frac{\bo_{max} }{|\bo_m|}}
 + ....\right)  = \BP_0 e^{\log{\bo_{max}/|\bo_m|}} = \bP_0 \frac{\bo_{max}}{|\bo_m|}
 \label{su_2}
\eeq
We see that the pairing susceptibility increases as $\bo_m$ decreases, but  remains finite and positive for all finite $\bo_m$, even when $\bO_{tot} =0$.  Re-doing calculations at a finite $\bO_{tot}$ we find the same result as in (\ref{su_2}), with $|\bo_m|$ replaced by
 max$(|\bo_m|, {\bar \bO}_{tot})$.  This implies  that  for $\gamma >0$, there is
 no signature of a pairing instability within the logarithmic approximation. The absence of a pairing instability within the logarithmic approximation generally implies that pairing does not develop at weak coupling and, if exists, is a threshold phenomenon.  In our case, the dimensionless coupling in the series in Eq. (\ref{su_2}) is a number equal to one, i.e., the problem we consider is not weak-coupling.   A weak-coupling limit can be reached if we extend the model and make the pairing interaction parametrically  smaller than the one in the particle-hole channel.  In practice, this is done by multiplying the interaction in the pairing channel by $1/N$, where $N >1$
  (Refs. \cite{raghu_15,Wang2016,Abanov_19,*Wu_19_1,paper_1,torroba_last}),  while the interaction in the particle-hole channel is left intact~
  \footnote{The extension to $N >1$ was originally rationalized by extending the original model to matrix $SU(N)$, with integer $N$, hence the notation. We treat $N$ as a continuous parameter.}
    In the extended model, $\log{\bo_{max}/|\bo_m|}$ gets multiplied by $1/N$, and at large $N$ the problem becomes weak-coupling.
 We show below that indeed  there is no pairing instability at large $N$.

  The case $\gamma=0+$ falls in between BCS and $\gamma >0$ cases.  Namely, we show below that
  the series of logarithms are not geometric, like the ones for $\gamma >0$, however by summing up the series  one does find the scale of a pairing instability, like in BCS. We assume and than verify that for relevant $\bo'_m$,  $\lambda \log{(\bL/|\bo'_m|)}$ is small for $\lambda \ll 1$, and neglect
  this term in the denominator of (\ref{ss_111_l_1}).
  We keep $\bO_{tot}$  non-zero to avoid divergencies and for simplicity set $\bo_m$ and $\bO_{tot}$ to be comparable. To logarithmic accuracy, we can then view $\bP$ as a function of a single parameter $\bO_{tot}$.

   Because the interaction is logarithmic, perturbation series hold in $\lambda \log^2 {(\bL/|\bo_m|)}$.
  Evaluating the pairing vertex in order-by-order calculations, we find (see Appendix for details)
  \beq
\BP (\bO_{tot}) = \BP_0 \left(1 +  \frac{1}{2} \lambda\log^2{\frac{\bL}{|\bO_{tot}|}} +  \frac{5}{24} \lambda^2 \log^4{\frac{\bL}{|\bO_{tot}|}} +  \frac{61}{720} \lambda^3 \log^6{\frac{\bL}{|\bO_{tot}|}} + ... \right)
 \label{su_2_1}
\eeq
At a first glance, the coefficients in (\ref{su_2_1}) are just some uncorrelated numbers. On a more careful look, we realize that
the series fall into
\beq
\BP (\bO_{tot}) = \BP_0 \frac{1}{\cos{\left(\sqrt{\lambda} \log{(\bL/|\bO_{tot}|)}\right)}}
  \label{su_2_2}
\eeq
Accordingly, the pairing susceptibility diverges when the argument of $\cos$ becomes $\pi/2$, i.e., at $\bO_{tot} = \bL e^{-\pi/2\sqrt{\lambda}}$. It is natural to associate this scale with $T_c$ (Ref. \cite{son}).  The susceptibility also diverges at a set of smaller $\bO_{tot,n} = \bL e^{-\pi(1+2n)/2\sqrt{\lambda}}$, but here we focus only on the  highest onset temperature.
 We see that $T_c$ has exponential dependence on the coupling, like in BCS theory, however the argument of the exponent contains $\sqrt{\lambda}$ rather than $\lambda$.   This in turn justifies the neglect of the self-energy, because for relevant frequencies,  $\lambda \log{(\bL/|\bo'_m|)} \sim \sqrt{\lambda} \ll 1$
  (the corrections due to self-energy have been analyzed in Ref. \cite{rischke}).

To summarize,  the case $\gamma = 0+$ is  similar to BCS in the  sense that  pairing occurs for arbitrary weak
  coupling ($T_c$ remains finite even if we replace $\lambda$ by $\lambda/N$ and set $N$ to be large).  However, for and finite $\gamma >0$, there is no indication of the pairing instability within the logarithmic approximation.

  \section{The differential equation for $\bP (\bo_m)$}
\label{sec:finite_gamma}

 We now analyze the linearized equation  for the pairing vertex for $\gamma >0$   beyond the logarithmic approximation.  To do this, we convert integral equation  (\ref{ss_111_l})
 into to a differential equation with certain boundary conditions and solve it
  non-perturbatively.

\subsection{The case $\bL = \infty$}

 We  first consider the case $\Lambda = \infty$.  We keep
 $N \geq 1$
as a parameter, as we need to verify our earlier  conjecture that there is no pairing instability for large enough $N$.
 To convert (\ref{ss_111_l}) into differential  equation, we use the fact that at small $\gamma$, the integral in the r.h.s. of (\ref{ss_111_l}) predominantly  comes from internal $\bo'_m$, which are either substantially  larger or substantially smaller than the external  $\bo_m$. We then split the integral over $\bo'_m$ into two parts:  in one we approximate $|\bo_m - \bo'_m|$ by $|\bo'_m|$, in the other by $|\bo_m|$.
 Introducing $z = |\bo_m|^\gamma$, we then simplify (\ref{ss_111_l}) to
    \beq
    \BP (z) = \frac{1}{N \gamma} \left[\int^{\infty}_{z} dy \frac{\BP (y)}{y (1+ \alpha y)} + \frac{1}{z} \int_0^z dy \frac{\BP (y)}{1+ \alpha y}\right]
   \label{su_6}
   \eeq
 where $\alpha = \gamma/\lambda$ ($\bo_{max} = (1/\alpha)^{1/\gamma}$).
Differentiating this equation twice over $z$ and replacing $\BP(z)$ by $\bD (z) = \alpha \bP (z) z/(1 + \alpha z)$, we obtain the  second order differential gap equation in the form (Refs. ~ \cite{Wang_H_17,Wang_H_18,Wang2016,paper_1})
   \beq
   (\BD_{{\text{diff}}} (z) (1+ \alpha z))^{''}  = - \left(1/4 - b_N^2\right) \frac{\BD_{{\text{diff}}} (z)}{z^2},
   \label{su_7}
   \eeq
   where  $(...)^{''} = d^2 (...)/dz^2$ and
   \beq
   b_N = \sqrt{\frac{1}{4}-\frac{1}{\gamma N}}
   \eeq
This $\BD_{{\text{diff}}} (z)$ has to be real and satisfy the boundary conditions
 at large $z$ and at $z=0$. At large $z$, we  expect perturbation theory to work, hence
 $\lim\limits_{z\rightarrow \infty }\BD_{{\text{diff}}} (z)=  \bP_0$.
The boundary condition at $z=0$ is set by the requirement that $\BD_{{\text{diff}}} (z)$ is
 normalizable, i.e., that the ground state  energy for a non-zero $\BD_{{\text{diff}}} (z)$ must be free from divergencies. This requirement imposes the condition that
$\int dz (\BD^2_{{\text{diff}}} (z)/z^2$ should be infra-red convergent (Refs. \cite{paper_1,Emil2020}).

\begin{figure}
	\begin{center}
		\includegraphics[width=0.49\columnwidth]{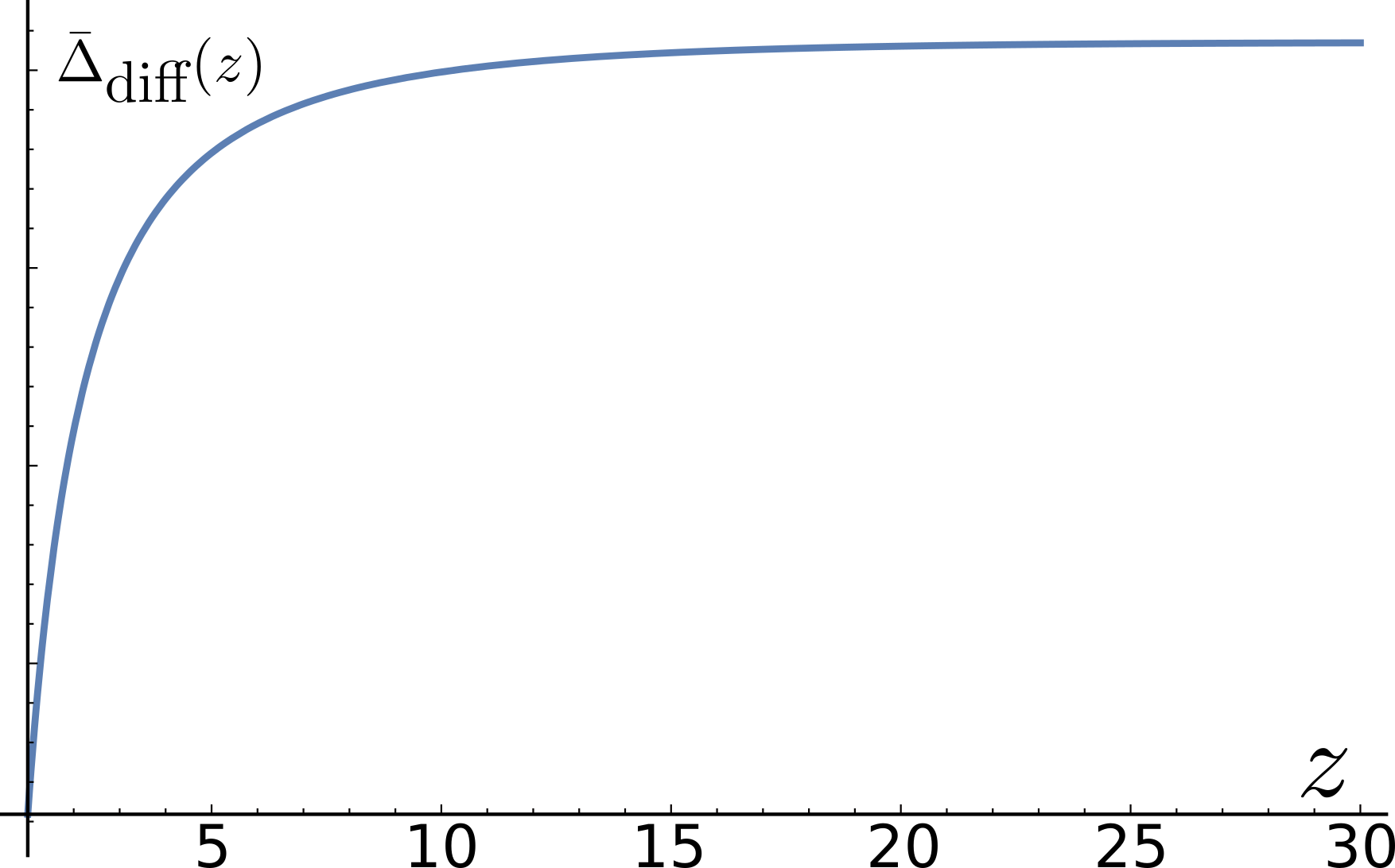}
		\includegraphics[width=0.49\columnwidth]{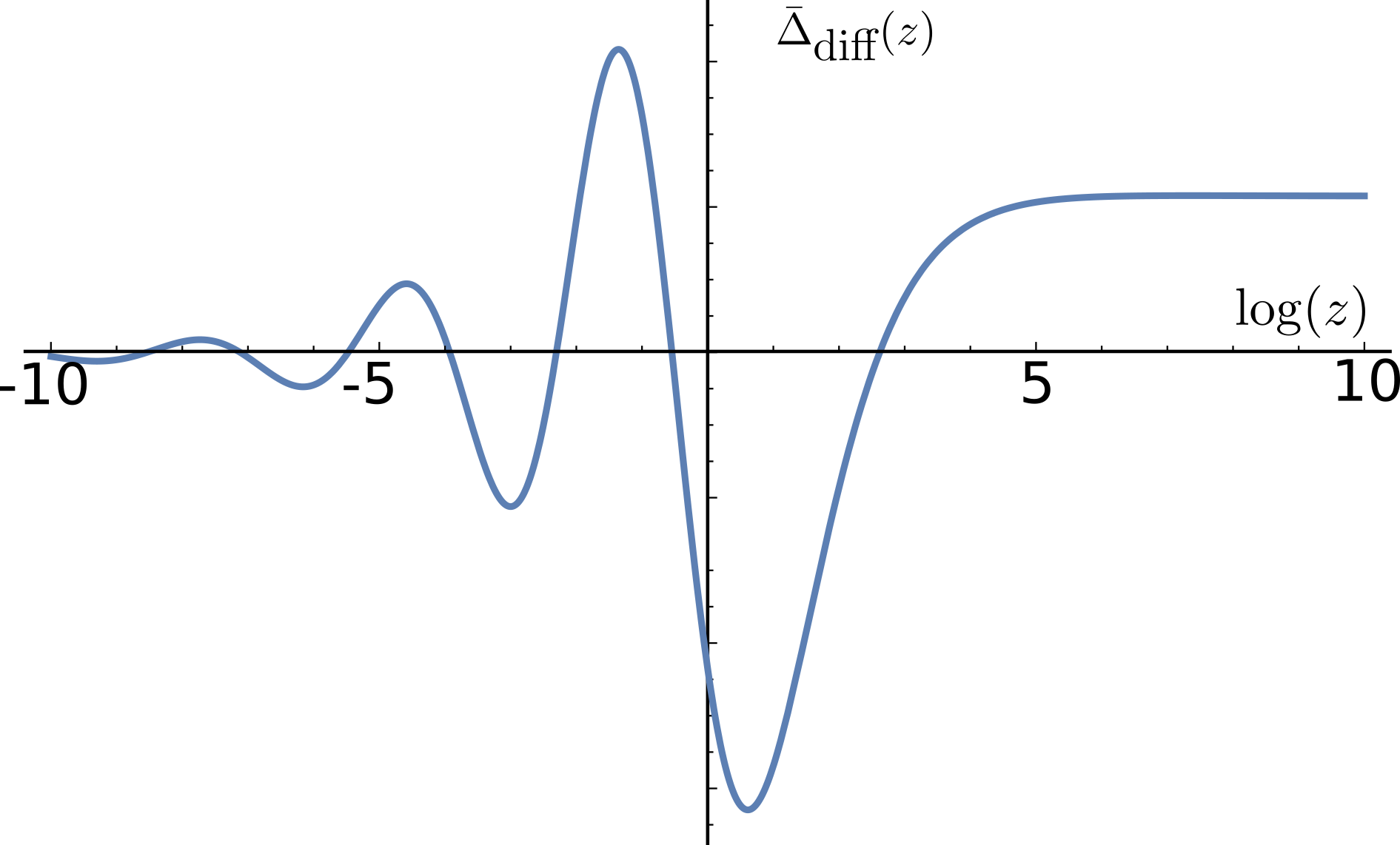}
		\caption{ Left: The function $\BD_{\text{diff}}(z)$, Eq. \eqref{eq:DeltaRealB}, for real $b_{N}=0.2$ ($N > N_{cr}$) and $\alpha =1$. Right: the same function for imaginary $b_{N}=2i$  ($N < N_{cr}$).  For real $b_N$ perturbation theory around the normal state is regular, which indicates that there is no pairing instability. For
 imaginary $b_N = i\beta_N$, $\BD_{\text{diff}}(z)$ oscillates at small $z$. We argue in the text that this implies that the normal state becomes unstable towards pairing.}
\label{fig:DeltaRealB}
	\end{center}
\end{figure}

The solution of (\ref{su_7}) is expressed via a Hypergeometric function, and the form of $\BD_{{\text{diff}}} (z)$
 depends on whether $b_N$ is real or imaginary, i.e., whether $N$ is larger or smaller than $N_{cr} =4/\gamma$.  When $N > N_{cr}$, $b_N$ is real, and the solution is
 \begin{widetext}
     \beq
\label{eq:DeltaRealB}
     \BD_{{\text{diff}}} (z) = C_1 S[b_N,z] + C_2 S[-b_N,z]
     \eeq
     where
     \beq
     S[b_{N},z]
= z^{b_N+1/2} {_2}F{_1} \left[\frac{1}{2} +b_N, \frac{3}{2} + b_N, 1 +2 b_N, -\alpha z\right].
 \label{su_9_3}
 \eeq
 \end{widetext}
  Both $S[b_N,z]$ and $S[-b_N,z]$ are sign-preserving functions.
 At small $z$,
$S[\pm b_N,z] \approx z^{1/2\pm b_N}$,
and at large $z$,  $S[\pm b_N,z]$ tend to finite values $\frac{4^{\pm b_{N}} \Gamma (1\pm b_{N})}{\sqrt{\pi } \Gamma \left(\frac{3}{2}\pm b_{N}\right)}$.
The boundary condition  at $z =0$  sets $C_2 =0$, and the one
  at $z \gg 1$ sets the linear relation between $C_1$ and $\bP_0$.
   We  plot  $\BD_{{\text{diff}}} (z)$
for $b_{N}=0.2$ and $\alpha =1$ in Fig. \ref{fig:DeltaRealB}.
  We see that $\BD_{{\text{diff}}} (z)$ remains positive for all $z$, i.e., the normal state remains stable with respect to pairing.    At $\gamma N \gg 1$, $b_N \approx 1/2-1/(\gamma N)$, and
  $\BD_{{\text{diff}}} (z)/\bP_0  \propto 1/|z|^{1/(N\gamma)}$.  This is the same result that we
   obtained by summing up the logarithms.  We see that in this parameter range the non-logarithmic terms just change the exponent from $1/N$ to $(\gamma/2) \sqrt{1-4/(N\gamma)}$.

 This behavior holds as long as $b_N$ is real, i.e.,
$N > N_{cr} =4/\gamma$.
For models with smaller $N < N_{cr}$, including our original model with $N=1$, $b_N$ is imaginary, $b_N= i \beta_N$, were
   $\beta_N = \sqrt{1/(N\gamma) - 1/4} = (1/N \gamma)^{1/2}  (1 - N/N_{cr})^{1/2}$.   In this situation,
  the
solution of the differential equation is
 \begin{widetext}
     \beq
 \BD_{{\text{diff}}} (z) = C \sqrt{z} \times Re\left(e^{i \psi} z^{i \beta_N}
  {_2}F{_1} \left[\frac{1}{2} + i\beta_N,  \frac{3}{2} + i\beta_N, 1 + 2  i \beta_N, -\alpha z\right] \right)
 \label{su_17}
 \eeq
 \end{widetext}
 The boundary condition at $z =0$ is satisfied  as at small $z$, $\bP_{{\text{diff}}} (z)$  acquires a form of a free quantum
particle with a "coordinate" $x = \log{z}$.
 At $z \gg 1$,  $\BD_{{\text{diff}}} (z)$ tends to a constant for a generic $\psi$, and the boundary condition
  sets up a linear relation between $C$ and $\bP_0$.

We plot $\BD_{{\text{diff}}} (z)$ in Fig. (\ref{fig:DeltaRealB}).
 We see that the gap function now  passes through a minimum
 at some $z^*$ and oscillates at smaller $z$.  The oscillating behavior at $z < z^*$
  could not be obtained within perturbation theory starting from $\bP (z) = \bP_0$ as the kernel in Eq. (\ref{ss_111_l}) is entirely positive.   This strongly suggests that the normal state is now unstable towards pairing, and  $(z^*)^{1/\gamma}$ sets the value of $\bT_c$.  To verify this, in Ref. ~\cite{paper_2}
   we solved the gap equation at a finite $T$. This calculation is a bit more tricky due to
    special behavior of fermions with Matsubara frequencies $\omega_m =\pm \pi T$, but nevertheless it confirms
     the two key points of our $T=0$ analysis that superconducting $\bT_c$ is finite and is generally of order $z^*$.

 Note that the value of $z^*$ depends on the magnitude of $\beta_N$, i.e., on the ratio $N_{cr}/N$.
   For small $\beta_N$, which holds when  $N$ is only slightly below $N_{cr}$,  the first sign change of
    $\chi_{pp} (z) \equiv \bar{\Delta }_{\text{diff}}/\bar{\Phi }_{0}$ occurs at small $z^* \sim e^{-a/\beta_N}$, where $a = O(1)$. For larger $\beta_N$,  the first sign change occurs at $\alpha z^* \sim 1/(N\gamma) \geq 1$, i.e., at $z^* \sim \lambda/N\gamma^2$, or, equivalently, at $\bo_m \sim \bo_{max} (1/(N \gamma)^{1/\gamma}$.
     This holds for the original model with $N=1$.

  We note in passing that $\BD_{{\text{diff}}} (z)$ contains two parameters: the overall factor $C$ and the phase factor $\psi$. The boundary condition at $z \gg 1$ then still leaves the freedom of, e.g., choosing $\psi$ for a given $C \sim \bP_0$.  This extra freedom comes about because it turns out~\cite{paper_1}
    that the homogeneous gap equation  (the one without $\bP_0 =0$) by itself has a non-zero solution at $T=0$,  such that $\BD_{{\text{diff}}} (z)$  is the sum of the induced and the  homogeneous solutions of (\ref{ss_111_l}). To demonstrate this more explicitly, we set $z >1/\alpha$, use the fact that in this range  a hypergeometric function reduces to a combination of Bessel and Neumann functions, and re-express (\ref{su_17}) as
\beq
\BD_{{\text{diff}}} (z)
\sim
 \frac{1}{\sqrt{z}} \left[A_1  J_1 \left(\sqrt{\frac{4 \beta^2_N+1}{\alpha z}}\right) + A_2 Y_1 \left(\sqrt{\frac{4\beta^2_N +1}{\alpha z}}\right)\right],
 \label{su_11}
 \eeq
 where $J_1$ and $Y_1$ are Bessel and Neumann functions, respectively.
 At small value of the argument, $J_1 (p) \approx p$ and $Y_1 (p) \approx 1/p$, i.e.,
 the first term vanishes at large $z$, while the second one tends to a constant. Then $A_2$ is uniquely determined  by the boundary condition set by $\bP_0$, while the $A_1$ piece is the solution of the equation without $\bP_0$.

    The existence of the solution of the linearized gap equation at $T=0$ (well below $T_c$) is a highly non-trivial feature of pairing at a QCP, that affects fluctuation corrections to superconducting order parameter.
  For our purposes, however, we only need to analyze only the induced solution to get an estimate of $T_c$. We can set  $A_1=0$ in (\ref{su_11}) and determine  $A_2$ from the
 boundary condition at $z \to \infty$.

  \subsection{The case of a finite $\bL$}

  At this stage, we have two different energy scales, which  we identified with $T_c$. Namely,  at a finite $\bL$ and $\gamma \to 0$, we found
   $\bT_c \sim \bL e^{-\pi/(2\sqrt{\lambda})}$.  At $\bL \to \infty$ and finite $\gamma$, we found $\bT_c \sim \bo_{max} (1/N\gamma)^{1/\gamma}$.   We now analyze the crossover between the two energies.  For definiteness,
    we consider the original model with $N=1$.

  We argued earlier that for the pairing at $\gamma = 0+$, fermions can be  treated as free quasiparticles, because for relevant fermions the ratio $\bS (\bo_m)/\bo_m \sim \sqrt{\lambda} \ll 1$.  The same holds for the case $\bL = \infty$ and $\gamma >0$. Here, the ratio $\bS (\bo_m)/\bo_m \sim 1/(\alpha z)$, and for $z \sim z^*$, $\bS (\bo_m)/\bo_m \sim \gamma \ll 1$. We now use this simplification and re-analyze the differential equation for $\bD (z)$   at a finite $\bL$.

\begin{figure}
\includegraphics[width=0.5\columnwidth]{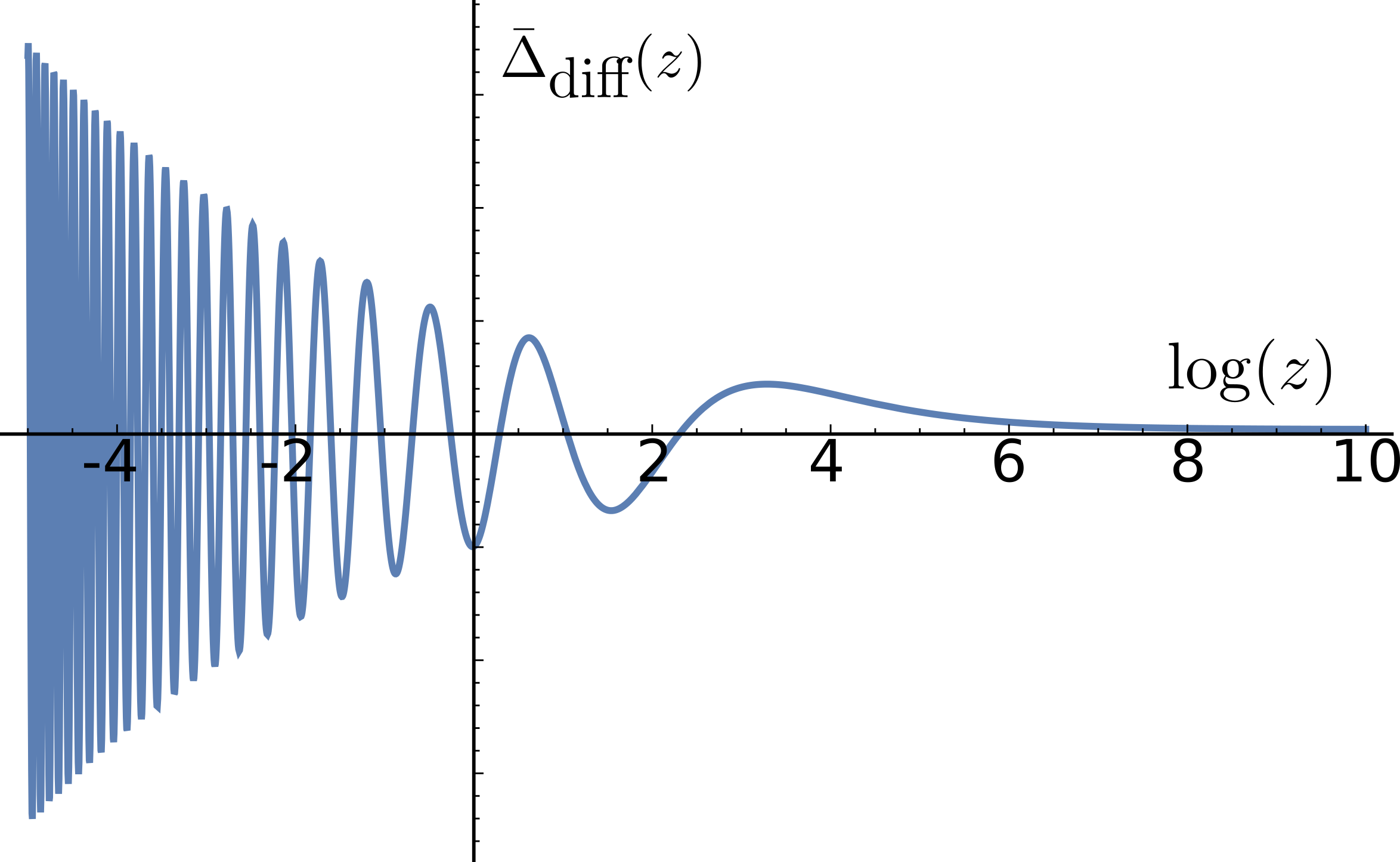}
\includegraphics[width=0.2\columnwidth]{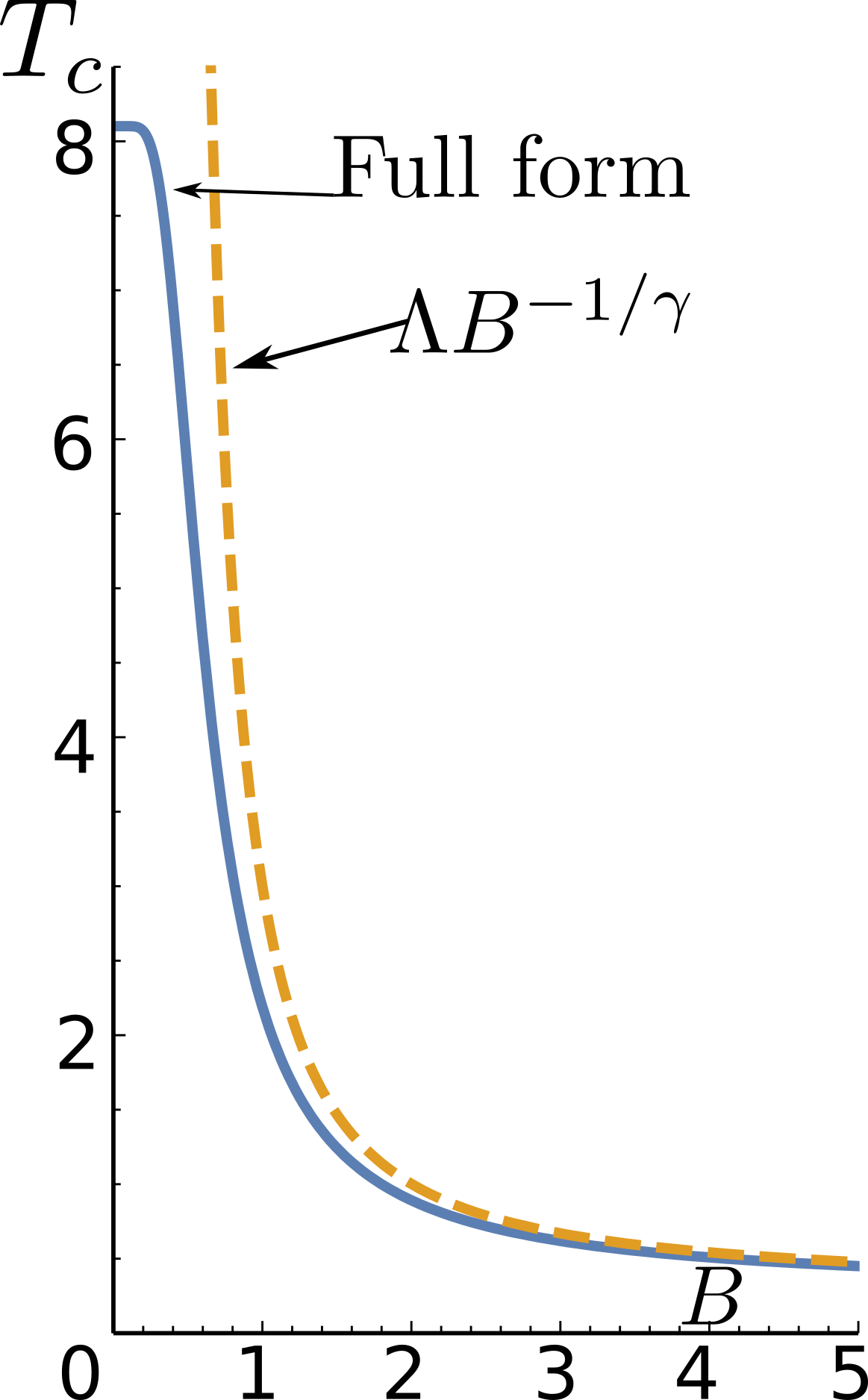}
\caption{Left: $\BD_{\text{diff}}(z)$ from \eqref{last_3a} for $B=0.3$, and $\bar{\Lambda}^{\ast} =10$. Right: the sketch of the evolution of $T_{c}$ as a function of $B$.}
\label{fig:DeltaLambda}
\end{figure}

  One can verify that the differential equation retains the same for as for  $\bL = \infty$:
   \beq
   (\BD_{{\text{diff}}} (z) z)^{''}  = - \frac{\lambda}{\gamma^2} \frac{\BD_{{\text{diff}}} (z)}{z^2},
   \label{su_10}
   \eeq
   and its solution is still a linear combination of Bessel and Neumann functions, Eq. (\ref{su_11}).
  However,  we have an extra  requirement
  \beq
  \BD_{{\text{diff}}} (\bL^*) =0,
\label{ya}
\eeq
where $\bL^* = (\bL)^\gamma$.
There is no solution of the homogeneous equation at a finite $\bL$, and Eq. (\ref{ya}) together
    with the  boundary condition  $\BD_{{\text{diff}}} (\infty) = \bP_0$  uniquely specify the coefficients $A_1$ and $A_2$ in (\ref{su_11}):
\bea
&&
\BD_{\text {diff}}
(z) = \frac{C}{\sqrt{z}} \times  \nonumber \\
&& \left[J_1 \left(\frac{2}{(\gamma \alpha z)^{1/2}}\right) Y_1 \left(\frac{2}{(\gamma \alpha \bL^*)^{1/2}}\right)
- Y_1 \left(\frac{2}{(\gamma \alpha z)^{1/2}}\right) J_1 \left(\frac{2}{(\gamma \alpha \bL^*)^{1/2}}\right)\right],
\label{last_3a}
\eea
 where $C \sim \bP_0$.
It is convenient to introduce a dimensionless parameter
\beq
B = \gamma \alpha \bL^* = \frac{\gamma^2}{\lambda} (\bL)^\gamma.
\eeq
In the left panel of Fig. \ref{fig:DeltaLambda} we plot $\BD_{\text{ diff}} (\bo_m)$
 for a representative $B =0.3$.  We see that the gap function is regular  at $|\bo_m| \leq \bL$, but passes through an extremum and oscillates at smaller $\bo_m$. The position of the first extremum depends on $B$.
 We now show that the two forms of $\bT_c$, which we found earlier, correspond to the limits $B \ll 1$ and $B \gg 1$.

 When $B \ll 1$, one can use the forms of Bessel and Neumann functions at large argument,
\bea
&& J_1 (x) \approx \sqrt{\frac{2}{\pi x}} \cos{(x -3\pi/4)}, \nonumber \\
&& Y_1 (x) \approx \sqrt{\frac{2}{\pi x}} \sin{(x -3\pi/4)},
\label{last_4a}
\eea
and obtain
\beq
\BD_{\text{ diff}} (z) \propto \frac{\bP_0}{z^{1/4}} \sin{\left(\frac{2}{B^{1/2}} - \frac{2}{(\alpha \gamma z)^{1/2}}\right)} = \frac{C}{z^{1/4}} \sin{\frac{2}{B^{1/2}} \left(1 - \left(\frac{\bL^*}{z}\right)^{1/2}\right)}
\label{last_5a}
\eeq
In the original variables $\bo_m$ and $\Lambda$, this reduces, to the leading order in $\gamma$, to
\beq
\BD_{{\text {diff}}} (\bo_m) \propto \frac{\bP_0}{|\bo_m|^{\gamma/4}} \sin{\left(\sqrt{\lambda} \log{\frac{|\bo_m|}{\bL}}\right)}.
\label{last_6a}
\eeq
 Associating $\bT_c$ with the position of the first extremum of
$\BD_{\text{ diff}} (\bo_m)$, we find $\bT_c \sim \bL e^{-\pi/(2\sqrt{\lambda})}$. This coincides with the estimate of $\bT_c$ from the analysis of
 the pairing susceptibility
  as a function of the total frequency of two fermions in a pair (see Eq. (\ref{su_2_2})).

In the opposite limit $B \gg 1$, we use $J_1 (2/B^{1/2}) \sim 1/B^{1/2}$ and $Y_1 (2/B^{1/2}) \sim B^{1/2}$ and
keep only  $Y_1 (2/B^{1/2})$.  We then obtain  from (\ref{last_3a}) that the dependence on $\bL$ appears only in the overall factor:
 \beq
\BD_{\text {diff}} (z) \propto \frac{1}{\sqrt{z}} J_1 \left(\frac{2}{(\gamma \alpha z)^{1/2}}\right),
\label{last_7}
\eeq
or, in terms of $\bo_m$,
\beq
\BD_{\text {diff}} (\bo_m) \propto \frac{1}{|\bo_m|^{\gamma/2}} J_1 \left(\frac{2}{B^{1/2}} \left(\frac{\bL}{|\bo_m|}\right)^{\gamma/2}\right)
\eeq
The first extremum is now located at
 $\bo^*_m  = (z^*)^{1/\gamma} \sim \bT_c \sim \bL/B^{1/\gamma}$, i.e.,  $\bT_c \sim (\lambda/\gamma)^{1/\gamma} (1/\gamma)^{1/\gamma}$. This agrees with the analysis at $\bL = \infty$.

In the right panel in Fig. \ref{fig:DeltaLambda}  we sketch
the evolution of $\bT_c$ as a function of $B$. In terms of $B$,   $\bT_c$  scales as
 $\bL/B^{1/\gamma}$ for $B \gg 1$ (the limit of large $\bL$ and finite $\gamma$),  and saturates at $\Lambda e^{-\pi/2\sqrt{\lambda}}$ for $B \ll 1$ (the limit $\gamma = 0+$ and finite $\bL$). The crossover  between the two regimes occurs at $B \sim e^{\pi \gamma/(2\sqrt{\lambda})} \approx 1$.

\begin{figure}
\includegraphics[width=0.6\columnwidth]{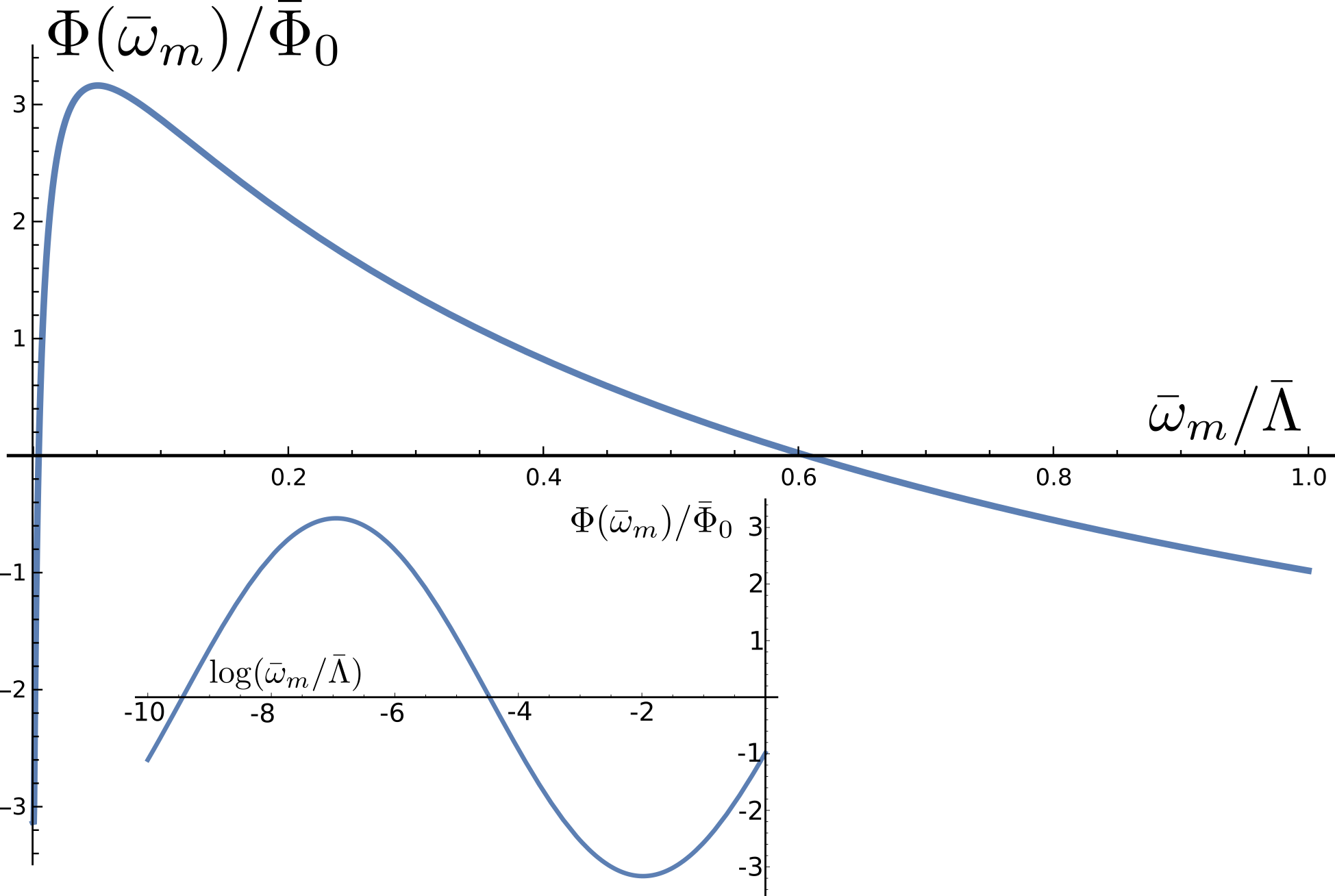}
\caption{\label{fig:phi_z_app_1}
 The function $\Phi (\bo_m)$ from Eq. (\ref{app1_4_a}). It passes through a maximum at $\bo_m \sim \bT_c \sim \bL e^{-\pi/(2\sqrt{\lambda})}$ and oscillates at smaller $\bo_m$ as a simple trigonometric function of $\log{\frac{\bL}{|\bo_m|}}$ (shown in the insert). }
   \end{figure}

 We note in passing that Eq. (\ref{last_6a}) could also be obtained directly, by converting the gap equation for $\gamma =0+$ into the differential equation~\cite{son2}. For this we depart from Eq. (\ref{ss_111_l_1}), neglect the self-energy term in the denominator, split the integral over $\omega'_m$ into
  contributions from $\omega'_m \gg \omega_m$ and $\omega'_m \ll \omega_m$, as we did in the derivation of (\ref{su_6}), and introduce logarithmic variables $x = \log{\Lambda/|\omega_m|} = \log{\bL/|\bo_m|}$, $x' = \log{\bL/|\bo'_m|}$. We then obtain
 \beq
 \Phi (x) = \lambda \left[\int_0^x  dx' x' \Phi(x') + x \int_x^\infty dx' \Phi (x') \right]
  \label{app1_2a}
 \eeq
 Differentiating twice over $x$ and introducing $y = \gamma x$, we obtain
 \beq
 \Phi^{''} (y) = -\frac{\lambda}{\gamma^2} \Phi (y)
   \label{app1_3}
 \eeq
 This equation is valid only for $0< y <\gamma^2/\lambda$, and the two boundary conditions are
 $\Phi (y=0) =\Phi_0$ and  $\Phi (y =\gamma^2/\lambda) =0$.
 The solution of (\ref{app1_3}), which satisfies the boundary conditions,  is, for small $\lambda$,
\beq
 \Phi (y) = -\bP_0 \frac{\sqrt{\lambda}}{\gamma} \sin{\left(\frac{\sqrt{\lambda}}{\gamma} \left(y - \frac{\gamma^2}{\lambda} \right)\right)}
   \label{app1_4}
 \eeq
 In the original units, this becomes
 \beq
 \Phi (\bo_m) = -\bP_0 \frac{\sqrt{\lambda}}{\gamma} \sin{\left(\sqrt{\lambda}\log{\frac{\bL}{|\bo_m|}} - \frac{\gamma}{\sqrt{\lambda}}\right)}
   \label{app1_4_a}
 \eeq
We plot $\Phi (\bo_m)$ in Fig. \ref{fig:phi_z_app_1}.
  We see that $\Phi (\bo_m)$ is sign-preserving  when $\log{\frac{\bL}{|\bo_m|}}$ is small, but
  oscillates when  $\log{\frac{\bL}{|\bo_m|}}$ gets larger.  Associating the position of the first extremum of $ \Phi (\bo_m)$ with $\bT_c$, as we did before, we find  the same $\bT_c \sim \bL e^{-\pi/(2\sqrt{\lambda})}$
   as at $B \ll 1$ in our earlier treatment.

 \section{RG analysis}
\label{sec:RG}

 The difference between   $\gamma =0+$ and at a finite $\gamma$ can be also understood by analyzing the flow of the running 4-fermion pairing vertex within the RG formalism.
  The RG equations for small $\gamma>0$ have been derived in Refs. ~\cite{max_last,Wang_H_17} and for $\gamma = 0+$ in Ref. \cite{son} (see also Refs. \cite{wilczek,pisarski}).

    The key point for the RG analysis is that at small  but finite $\gamma$, the interaction $V(\bO_m)$ in (\ref{a_1}) can be  approximated by
    \beq
    V(\bO_m) = \frac{\lambda}{|\bO_m|^\gamma} \log{\frac{\bL}{|\bO_m|}}
   \label{a_1_a}
   \eeq
 and both the logarithmic and the power-law dependence have to be kept.
  This interaction can be further re-expressed in terms of $L = \log{\frac{\bL}{|\bO_m|}}$ as
  the product of $L$ and the running effective ${\tilde V} (L) = {\tilde V} (0) e^{\gamma L}$, where
  ${\tilde V} (0) = \lambda (1/\bL)^\gamma =\gamma^2/B$.

  The interaction $V(L) = {\tilde V} (L) L $ acts as  the source for  the running 4-fermion pairing vertex, which we label as $V_{SC} (L)$. Without the source, $V_{SC} (L)$ would obey a BCS RG equation  ${\dot V}_{SC} (L)  = V^2_{SC} (L)$. With the source, the equation becomes
    \beq
{\dot V}_{SC} (L)  = V^2_{SC} (L) + {\tilde V} (L)
\eeq
The solution of this equation, subject to $V_{SC} (0) =0$ (no pairing without the source), is
~\cite{max_last,Wang_H_17}
\beq
V_{SC} (L) = \frac{\gamma}{B^{1/2}} e^{\gamma L/2} \frac{J_1\left(\frac{2}{B^{1/2}}  e^{\gamma L/2}\right) Y_1 \left(\frac{2}{B^{1/2}}\right) - Y_1\left(\frac{2}{B^{1/2}} e^{\gamma L/2}\right) J_1 \left(
\frac{2}{B^{1/2}}\right)}{J_0\left(\frac{2}{B^{1/2}} e^{\gamma L/2}\right) Y_1 \left(
\frac{2}{B^{1/2}}\right) - Y_0\left(\frac{2}{B^{1/2}} e^{\gamma L/2}\right) J_1 \left(\frac{2}{B^{1/2}}\right)}
\label{last_9}
\eeq
For $B \ll 1$,  relevant values of the arguments of Bessel and Neumann functions are large, and using
  using their asymptotic forms, Eq. (\ref{last_4a}),
 we find that, to leading order in $\gamma$,
 \beq
 V_{SC} (L) \propto \tan{ (\sqrt{\lambda} L)}
 \label{last_10}
 \eeq
 The 4-fermion vertex diverges at the same $\bo^*_m = \Lambda e^{-\pi/(2 \sqrt{\lambda})}$ that we obtained before.

 In the opposite limit $B \ll 1$, we use that  $Y_1 (2/B^{1/2}) \gg J_1(2/B^{1/2})$. Keeping
 only $Y_1 (2/B^{1/2})$ in (\ref{last_9}), we obtain
  \beq
  V_{SC} (L) = \frac{\gamma}{B^{1/2}} e^{\gamma L/2}
    \frac{J_1\left(\frac{2}{B^{1/2}} e^{\gamma L/2}\right)}{J_0\left(\frac{2}{B^{1/2}} e^{\gamma L/2}\right)}
\label{last_12}
 \eeq
 The 4-fermion vertex now diverges at the first zero of  $J_0 (p)$, i.e., at
 \beq
  \frac{2}{B^{1/2}} \left(\frac{\bL}{|\bo_m|}\right)^{\gamma/2} = \frac{2}{(\gamma  \lambda z)^{1/2}} \approx 2.4
 \eeq
  Solving for $\bo_m$, we obtain $\bo^*_m = (z^*)^{1/\gamma} \sim \bo_{max} (1/\gamma)^{1/\gamma}$. This
   agrees the result of our  earlier analysis of the case $\bL \to \infty$ and $\gamma >0$.

\section{Conclusions}
\label{sec:conclusions}

The goal of this work was to analyze the crossover from  a conventional BCS pairing by a massive boson to
a pairing  at a quantum-critical point towards
  some  particle-hole order, when the pairing boson becomes massless.
     We considered a subset of quantum-critical models, in which the  pairing boson is a slow mode compared to fermions, and an effective low-energy theory is purely dynamical, with an effective dynamical
       interaction
     $V(\Omega_m) \propto 1/|\Omega_m|^\gamma$, up to some upper cutoff $\Lambda$.
      The case $\gamma =0$ corresponds to BCS theory od pairing from a Fermi liquid.
       We considered the pairing at a small, but finite
       $\gamma$, when the normal state is a NFL,  and the  case $\gamma = 0+$, when $V(\Omega_m) \propto \log{\Lambda/|\Omega_m|}$, and the normal state is marginal Fermi liquid.
        We demonstrated that for $\gamma = 0+$, the pairing instability can still be detected by summing up
         series of Cooper logarithms, but for a finite $\gamma$ one needs to go beyond the leading logarithmic approximation to analyze the pairing.  We argued that in this last case, the pairing occurs only if the interaction exceeds some finite threshold.  We approximated the original integer gap equation by
          the differential one and solved it. This allowed us to
         identify the threshold at $\gamma >0$ and the frequency scale, associated with superconductivity, once the interaction exceeds the threshold,  We found the crossover between the pairing at a finite $\gamma$ and at
         $\gamma =0+$ and identified the parameter responsible for the crossover.  We obtained the same crossover by analyzing the non-BCS RG equation for the running 4-fermion pairing vertex.

  \acknowledgements
 We thank   B. Altshuler,  A. Finkelstein,  S. Karchu, S. Kivelson, I. Mazin, M. Metlitski,  V. Pokrovsky,   S. Raghu,  S. Sachdev,  T. Senthil, J. Schmalian, D. Son,  G. Torroba,  E. Yuzbashyan, C. Varma,  Y. Wang, Y. Wu, and S-S Zhang for useful discussions.   The work by  AVC was supported by the NSF DMR-1834856.  AVC acknowledges the hospitality of KITP at UCSB, where part of the work has been conducted. The research at KITP is supported by the National Science Foundation under Grant No. NSF PHY-1748958.

 \appendix
 \section{Derivation of Eq. (\ref {su_2_1})}

\begin{figure}
	\begin{center}
\includegraphics[width=0.6\columnwidth]{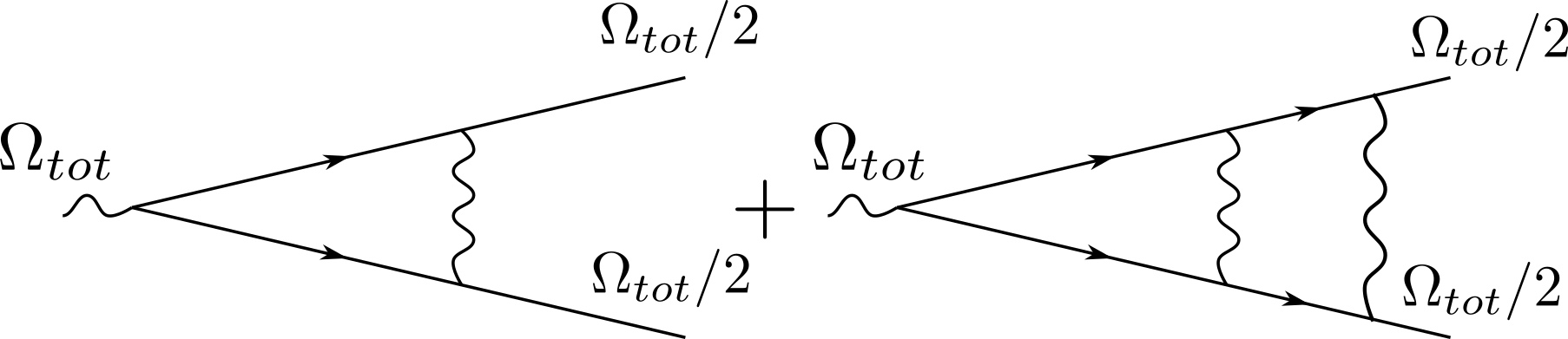}
		\caption{Diagrammatic representation of one-loop and two-loop renormalization of the pairing vertex $\Phi_0$ for the case $\gamma = 0+$.  Solid lines are propagators of free fermions,  wavy lines are $V(\bO_m) = \lambda \log{\bL/|\bO_m|}$, and the two-fermion vertices are $\Phi_0$.  For definiteness, we set  the frequencies of external fermions to be $\Omega_{tot}/2$.
}
\label{fig:OneLoopTwoLoop}
	\end{center}
\end{figure}
In this Appendix we present some details of the derivation of order-by-order logarithmic renormalization of the pairing vertex $\BP (\bO_{tot})$ for the case $\gamma = 0+$.  The Eliashberg equation for
$\BP (\bO_{tot})$  is Eq. (\ref{ss_111_l_1}). Adding a constant $\bP_0$ to the r.h.s. and expanding in powers of the coupling $\lambda$, we obtain the series in $\lambda \log^2{\frac{\bL}{|\bO_{tot}|}}$, where, we remind, $\bL$ in the upper cutoff, and all quantities are in units of our base energy ${\bar g}$.
 As we discussed in the main text, fermionic self-energy (the log term in the denominator of (\ref{ss_111_l_1})) is small at relevant $\log {\frac{\bL}{|\bO_{tot}|}} \sim 1/\sqrt{\lambda}$ and can be safely neglected.

We show the calculations at one-loop and two-loop order. The corresponding renormalizations are graphically presented in Fig. \ref{fig:OneLoopTwoLoop}. In this Figure, the two-fermion vertex is $\bP_0$, the wavy line is $V(\bO_m) = \lambda \log {\frac{\bL}{|\bO_m|}}$, and we set both external frequencies to be $\bO_{tot}/2$.     The one-loop result is obtained by integrating over a single internal frequency and is
\beq
\lambda \int_{|\bO_{tot}|}^{\bL} \frac{d \bo_m}{\bo_m} \log{\frac{\bL}{|\bo_m|}} = \frac{\lambda}{2} \left(\log\frac{\bL}{|\bO_{tot}|}\right)^2
\eeq

At two-loop order we  have to integrate over two internal frequencies $\bo_m$ and $\bo'_m$. We introduce $\bo_m = \bO_{tot} x$, $\bo'_m = \bO_{tot} y$ and use the fact that the leading logarithm comes from $x, y \gg 1$.
 Taking this limit, we obtain the two-loop  correction to $\bP_0$ in the form
\beq
\lambda^2 \int_{\sim 1}^{\bL/|\bO_{tot}|}  \frac{dx}{x}\int_{\sim 1}^{\bL/|\bO_{tot}|}
\frac{dy}{y} \left[\left(\log\frac{\bL}{|\bO_{tot}|} + \log{\frac{1}{x}}\right) \left(\log\frac{\bL}{|\bO_{tot}|} + \frac{1}{2} \log{\frac{1}{|x-y|}} + \frac{1}{2} \log{\frac{1}{x+y}} \right)\right]
\eeq
A simple analysis shows that the  highest power of $\log\frac{\bL}{|\bO_{tot}|}$  comes from the ranges $x \gg y$ and $y \gg x$.  Evaluating the contributions from these regions, we obtain
\bea
&& \int_{\sim 1}^{\bL/|\bO_{tot}|}  \frac{dx}{x}\int_{\sim 1}^{\bL/|\bO_{tot}|}
\frac{dy}{y} \left[ \log{\frac{1}{x}} \left(\frac{1}{2} \log{\frac{1}{|x-y|}} + \frac{1}{2} \log{\frac{1}{x+y}}\right) \right] = \frac{3}{8} \left(\log\frac{\bL}{|\bO_{tot}|}\right)^4 \nonumber \\
&&\frac{1}{2} \log\frac{\bL}{|\bO_{tot}|} \times
\int_{\sim 1}^{\bL/|\bO_{tot}|}  \frac{dx}{x}\int_{\sim 1}^{\bL/|\bO_{tot}|}
\frac{dy}{y} \left(\log{\frac{1}{|x-y|}} +  \log{\frac{1}{x+y}}\right) = - \frac{2}{3} \left(\log\frac{\bL}{|\bO_{tot}|}\right)^4 \nonumber \\
&&  \log\frac{\bL}{|\bO_{tot}|} \times
\int_{\sim 1}^{\bL/|\bO_{tot}|}  \frac{dx}{x}\int_{\sim 1}^{\bL/|\bO_{tot}|}
\frac{dy}{y} \log{\frac{1}{x}} = - \frac{1}{2} \left(\log\frac{\bL}{|\bO_{tot}|}\right)^4
\eea
Collecting all contributions, we obtain for the two-loop renormalization
\beq
 A \lambda^2 \left(\log\frac{\bL}{|\bO_{tot}|}\right)^4, ~~ A =1-\frac{1}{2} - \frac{2}{3} + \frac{3}{8} = \frac{5}{24}
 \eeq
 This leads to Eq. (\ref{su_2_1}) in the main text.

\bibliography{dzyal_90}

\end{document}